\documentclass[aps,pra,10pt,twocolumn,amsmath,amssymb]{revtex4-2}

\usepackage[utf8]{inputenc}
\usepackage{bm,bbold}
\usepackage{graphicx}
\usepackage[colorlinks=true,citecolor=blue,breaklinks=true,linkcolor=blue,linktocpage=true,urlcolor=blue,pagebackref=false]{hyperref}

\newcommand{\llangle}{\langle\!\langle}
\newcommand{\rrangle}{\rangle\!\rangle}

\begin{document}
\title{Quantum dissipation at conical intersections of quasienergies}

\author{Sigmund Kohler}
\affiliation{Instituto de Ciencia de Materiales de Madrid, CSIC, E-28049 Madrid, Spain}

\date{\today}

\begin{abstract}
We investigate the properties of Floquet states in the vicinity of a
conical intersection of quasienergies and work out the consequences of the
underlying spatio-temporal symmetries for a driven two-level system coupled
to an ohmic heat bath. We find that on manifolds with constant quasienergy
splitting, the mean energies of the Floquet states are continuously interchanged. In the
presence of dissipation, the parameter dependence of the stationary
populations generally resembles that of the mean energies.  In turn, the
mean energies are an indicator for the qualitative behavior of the density
operator in the long-time limit.  A further consequence of the symmetries
is that for specific driving parameters, the stationary state may be fully
mixed even at arbitrarily low temperatures.  For large driving frequencies,
such states with maximal entropy are found in the whole vicinity of the
intersection, which can be explained by a chirality emerging in this limit.
Analytical results beyond a high-frequency approximation are illustrated by
numerical data.
\end{abstract}

\maketitle

\section{Introduction}
Floquet engineering is the construction of effective Hamiltonians by
acting with ac fields or periodic pulses upon a quantum system.  The idea
originates in the suppression of tunneling in bistable systems
\cite{GrossmannPRL91} and tight-binding models \cite{HolthausPRL92}.  It
has been proposed also for many-particle states in double quantum dots
\cite{CreffieldPRB02a}, cold atoms \cite{GongPRL09, EckardtRMP17}, and
photonics \cite{LonghiPRA11}.  Experimental realizations have been achieved
with quantum dots \cite{StehlikPRB12, ForsterPRL14, ChenPRB21},
superconducting qubits \cite{SillanpaaPRL06, BernsNL08}, and optical
lattices \cite{LignierPRL07}.  These ideas have been extended to tuning
topological properties \cite{GomezLeonPRL13, AtalaNP13} and to artificial
gauge fields \cite{EckardtRMP17, AidelsburgerCRP18}.

A theoretical tool for studying ac-driven systems is Floquet theory with its
quasienergies \cite{ShirleyPR65, SambePRA73} which emerge
in the formal expression for the propagator and, thus, provide much
insight.  For example, suppression of tunneling can be attributed to
exact quasienergy crossings \cite{GrossmannPRL91, HolthausPRL92} which
require a spatio-temporal symmetry known as generalized parity
\cite{PeresPRL91, GrossmannEL92}.
When a small perturbation such as a static detuning breaks this symmetry,
the exact quasienergy crossings become avoided.  Then, as will be shown,
the quasienergy surfaces as a function of any original parameter and the
detuning form a double cone with a degeneracy at the tip, a so-called
conical or diabolical intersection.  Conical intersections have been
investigated mainly in the rather different context of breakdown of
adiabaticity in Born-Oppenheimer theory, where both coherent
\cite{YarkonyRMP96, Domcke2011, Larson2020} and dissipative \cite{ChenFD16,
DuanJPCL16} time-evolution have been considered.

Dissipative transitions between Floquet states lack detailed balance
\cite{KohlerPR05}.  Therefore, the stationary state generally is not a
canonical ensemble, but must be computed with a quantum master equation
\cite{BlumelPRA91, KohlerPRE97}.  Even for minimal dissipation, the
populations may significantly depend on the operator that couples the
system to the environment \cite{FerronPRL12, BlattmannPRA15,
EngelhardtPRL19}.  Only in the high-frequency
limit, one may find Floquet-Gibbs states, which are canonical states with
eigenenergies replaced by quasienergies \cite{ShiraiPRE15, ShiraiNJP16}.
In some cases, by contrast, the mean energies seem
more relevant for the populations \cite{KohlerPRE98, KetzmerickPRE10}.
For many of these results, level crossings play an important
role \cite{KohlerPRE98, FerronPRL12, BlattmannPRA15, EngelhardtPRL19,
ChenPRB21, IvakhnenkoPR23}.  In particular, it has been found
\cite{EngelhardtPRL19} that in the absence of a static detuning, at exact
quasienergy crossings, the populations of Floquet states may be
discontinuous.

Here, we extend previous work \cite{EngelhardtPRL19} on the driven
two-level system by considering the full conical intersection, i.e., by
investigating the impact of an additional detuning.  It will turn
out that sufficiently close to the cone tip, the physics is widely
determined by the symmetry operation responsible for the emergence of the
crossing.  Moreover, the analysis sheds light on the role of the mean
energies for dissipative transitions between Floquet states.

In Sec.~\ref{sec:coherent} we derive generic properties of the
Floquet states in that region without resorting to a high-frequency limit.
Section~\ref{sec:dissipative} is devoted to quantum dissipation, in
particular to the populations of the Floquet states in the long-time limit,
while conclusions are drawn in Sec.~\ref{sec:conclusions}.

\section{Floquet states at conical intersections}
\label{sec:coherent}

As a model, we consider the ac-driven two-level Hamiltonian
\begin{equation}
H(t) = \frac{\Delta}{2}\sigma_x + \frac{1}{2} [\varepsilon + A\cos(\Omega
t)] \sigma_z ,
\label{H}
\end{equation}
with tunneling $\Delta$, static detuning $\varepsilon$, driving amplitude $A$
and frequency $\Omega$, while $\sigma_x$ and $\sigma_z$ denote Pauli matrices.

Owing to time periodicity, the corresponding Schr\"odinger equation
possesses a complete set of solutions of the form $e^{-iqt}
|\phi(t)\rangle$ with quasienergy $q$ and Floquet state
$|\phi(t)\rangle$ which shares the time periodicity of the driving
\cite{ShirleyPR65, SambePRA73}.  Floquet states are eigensolutions of the operator
$\mathcal{H} = H(t)-i\partial_t$ (in units with $\hbar=1$).
Quasienergies come in a Brillouin zone structure, i.e., for any
$|\phi(t)\rangle$ and integer $n$, $|\phi^{(n)}(t)\rangle = e^{-in\Omega
t}|\phi(t)\rangle$ is a physically equivalent Floquet state with
quasienergy $q+n\Omega$.  To characterize a Floquet state, one
may employ its mean energy $E$ which is the expectation value of
$H(t)$ averaged over the driving period. Equivalent states possess
equal mean energy, as is expected for an observable.
For the present purpose, it is convenient to consider Floquet states as
elements of Sambe space which is the direct product of the underlying
Hilbert space and the space of $2\pi/\Omega$-periodic functions
\cite{SambePRA73}. We denote its elements by a double angle, $|\phi\rrangle$,
where $|\phi(t)\rangle = \langle t|\phi\rrangle$ and $E = q +
\llangle\phi| i\partial_t|\phi\rrangle$.

Hamiltonian \eqref{H} is invariant under time-reversal $t\to -t$ which
allows one to write the Floquet state $|\phi(t)\rangle$ as Fourier series
with real coefficients.  Moreover, the chirality $C=i\sigma_y K$ with $K$
denoting complex conjugation maps $\mathcal{H} \to -\mathcal{H}$, which
implies that non-zero quasienergies come in pairs with opposite sign.
Without loss of generality, we will choose the Floquet states from the
Brillouin zone symmetric around zero, $-\Omega/2 \leq q < \Omega/2$, and
label them such that $q_0\leq 0$, $q_1 = -q_0$.
Notice, however, that this order is arbitrary and by itself cannot
justify designations such as ``ground state''.

\begin{figure}
\centerline{\includegraphics{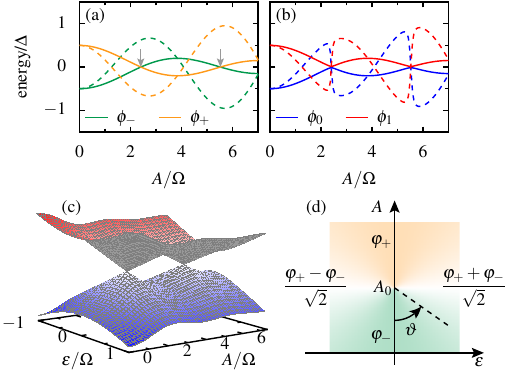}}
\caption{(a) Quasienergies $q_\pm$ (solid lines) and mean energies $E_\pm$
(dashed) for driving frequency $\Omega=25\Delta$ in the absence of a static
detuning, such that the Floquet states possess definite
generalized parity indicated by colors (green: even, orange: odd). We focus
on the vicinity of the quasienergy crossings marked by grey arrows.
(b) The same for static detuning $\varepsilon=0.03\Delta$, i.e., in the
absence of generalized parity. The color code now refers to the sign of the
quasienergy.
(c) Quasienergies $q_0$ and $q_1$ as a function of the detuning
$\varepsilon$ and the driving amplitude $A$. 
(d) Floquet state with lower quasienergy, $|\phi_0\rrangle$, in the
vicinity of a degeneracy point $(\varepsilon,A)=(0,A_0)$ in terms of the
basis states $|\varphi_\pm\rrangle$. The underlying color indicates the
corresponding mean energy $E_0 = E_-\cos\vartheta$ [green: $E_0>0$, cf.\
panel (a)].}
\label{fig:spectrum}
\end{figure}

A $\mathbb{Z}_2$ symmetry with direct consequences for the physics discussed herein
is the generalized parity $G = \sigma_x \exp(\pi\partial_t/\Omega)$ which consists
of a unitary transformation with $\sigma_x$ in Hilbert space and a time shift by
half a period.  It is present for $\varepsilon=0$.  Then, the
Floquet states can be classified as even or odd depending on the respective
eigenvalue $\pm 1$ of $G$ \cite{GrossmannEL92, PeresPRL91}.  Quasienergies of
states with different generalized parity may form exact crossings with
accidental degeneracies.  Also the mean energies may
exhibit exact crossings which, however, need not coincide with the
quasienergy crossings, see Fig.~\ref{fig:spectrum}(a).
Two properties of the generalized parity will turn out important.  First,
$\sigma_x$ and $i\partial_t$ are even under transformation with $G$, while
$\sigma_z$ is odd.  Second, from the above definition of the equivalent
states in the time domain, it is obvious that for a Floquet state with
generalized parity $\pm1$, the equivalent state $|\phi^{(n)}\rrangle$ has
generalized parity $\pm(-1)^n$.

The situation changes in the presence of a detuning $\varepsilon\neq 0$.
Then the Floquet states no longer possess generalized
parity, and exact quasienergy crossings turn into avoided crossings, as can
be witnessed in Fig.~\ref{fig:spectrum}(b).  This implies that the
quasienergies are degenerate only at isolated points of parameter space
$(\varepsilon,A)$.  Therefore, they form accidental symmetry-allowed
conical intersections, as is shown in Fig.~\ref{fig:spectrum}(c).
Henceforth, we focus on the vicinities of such intersections.

We employ degenerate perturbation theory using the
Floquet states $|\varphi_\pm\rrangle = \pm G|\varphi_\pm\rrangle$ for $H_0(t)
\equiv \frac{1}{2}\Delta\sigma_x + \frac{1}{2}A_0\sigma_z\cos(\Omega t)$ as
zeroth order, where $A_0$ denotes the driving amplitude at the respective
quasienergy crossing (the notation $\varphi_\pm$ is restricted to
$\varepsilon=0$ and $A=A_0$).  In the basis chosen, $i\partial_t = G^\dagger
i\partial_t G$ and $\sigma_x = G^\dagger \sigma_x G$ are diagonal, while
$\sigma_z = -G^\dagger \sigma_z G$ is off-diagonal.
For large $\Omega$, the amplitude at the crossing, $A_0$, can be expressed
with a root of a Bessel function \cite{GrossmannEL92}, but the present
investigation is not restricted to this limit.  Let us emphasize that while
the $|\varphi_\pm(t)\rangle$ form a basis of the Hilbert space, the full
Sambe space is spanned by all $|\varphi_\pm^{(n)}\rrangle$.
By evaluating the matrix elements of $H_1(t) = H(t)-H_0(t)$ in the subspace
considered, one finds the time-independent perturbation $H_1' =
\frac{1}{2}(a\sigma_z' + b\sigma_x')$ with
\begin{align}
a ={}& \frac{A_0-A}{A_0}
	\big(2E_- + \Delta\llangle\varphi_-|\sigma_x|\varphi_-\rrangle \big)
\equiv r\cos\vartheta ,
\label{a}
\\
b ={}& \varepsilon\llangle\varphi_-|\sigma_z|\varphi_+\rrangle
\equiv r\sin\vartheta ,
\label{b}
\end{align}
where $r=\sqrt{a^2+b^2}$ measures the distance to the cone tip.  Hence, the
perturbation theory holds for $r \ll \Omega,\varepsilon$.  Unlike in the
usual high-frequency approximation (see the Appendix), the present approach
does not require $\Delta\ll\Omega$ as long as $|A-A_0|\ll A_0$.  The prime
refers to the time-dependent basis given by the states
$|\varphi_\pm(t)\rangle$ with the mean energies $E_\pm =
\llangle\varphi_\pm|H(t)|\varphi_\pm\rrangle$.  To obtain the expression
for $a$, we have written the perturbation $(A-A_0)\sigma_z\cos(\Omega t)$
in terms of $H_0(t)-i\partial_t$ whose eigenvalues are the quasienergies
$q_\pm$ which for $A=A_0$ vanish according to the definition of $A_0$.
Moreover, we have used the already mentioned relation $E_\pm = q_\pm
+\llangle\varphi_\pm| i\partial_t |\varphi_\pm\rrangle$.

Diagonalization of $H_1'$ provides the perturbed Floquet states
\begin{align}
|\phi_0\rrangle &={}
	\cos(\vartheta/2)|\varphi_-\rrangle
	+ \sin(\vartheta/2)|\varphi_+\rrangle
\label{phi0}
\\
|\phi_1\rrangle &={}
	- \sin(\vartheta/2)|\varphi_-\rrangle
	+ \cos(\vartheta/2)|\varphi_+\rrangle
\label{phi1}
\end{align}
and their quasienergies $q_{0,1} = \pm r/2$.  Since the off-diagonal matrix
element
$\llangle\varphi_+| i\partial_t |\varphi_-\rrangle = 0$, the corresponding
mean energies read
\begin{equation}
E_0 = E_-\cos\vartheta = -E_1.
\label{E0}
\end{equation}
Figure \ref{fig:spectrum}(d) visualizes the definition of the angle
$\vartheta$ and the result of the perturbation theory.
Interestingly, values of $\vartheta$ that differ by $\pi$ give rise to the
following duality.  Upon the shift $\vartheta\to\vartheta+\pi$,
$|\phi_0(t)\rangle$ becomes proportional to $|\phi_1(t)\rangle$ and vice versa.
However, while the shape of the Floquet states is interchanged,
their quasienergies remain the same, i.e., $q_0<0$ and $q_1>1$ for all values of
$\vartheta$.

A central ingredient of our analysis will be the action of the generalized
parity operator $G$ on the Floquet states and their equivalent states.  Using
$G|\varphi_\pm^{(n)}\rrangle = \pm(-1)^n|\varphi_\pm^{(n)}\rrangle$, one
obtains from Eqs.~\eqref{phi0} and \eqref{phi1} the relations
\begin{align}
G|\phi_0^{(n)}\rrangle ={}& (-1)^n \left(
- |\phi_0^{(n)}\rrangle \cos\vartheta
+ |\phi_1^{(n)}\rrangle \sin\vartheta
\right) ,  \label{Gphi0}
\\
G|\phi_1^{(n)}\rrangle ={}& (-1)^n \left(
|\phi_0^{(n)}\rrangle \sin\vartheta
+ |\phi_1^{(n)}\rrangle \cos\vartheta
\right) .  \label{Gphi1}
\end{align}
For the particular value $\vartheta = \pm\pi/2$, $G$ maps the one Floquet
state to the other, i.e.\ $G|\phi_0^{(n)}\rrangle \propto
|\phi_1^{(n)}\rrangle$ and $G|\phi_1^{(n)}\rrangle \propto
|\phi_0^{(n)}\rrangle$.
Henceforth, we use the relations derived in this section to draw
conclusions on the stationary state established by the coupling to a heat
bath.

\section{Dissipation and stationary state}
\label{sec:dissipative}

To describe quantum dissipation, we couple the driven system to a heat bath
modeled as harmonic oscillators with the Hamiltonians \cite{LeggettRMP87,
HanggiRMP90, Weiss2012, LandiRMP22} $H_\text{bath} = \sum_\nu \omega_\nu
a_\nu^\dagger a_\nu$ and $H_\text{int} = S\xi$, where $\xi = \sum_\nu
\lambda_\nu (a_\nu^\dagger+a_\nu)$ is a collective bath coordinate and $S$
a system coordinate to be specified later.  The influence of the bath is
determined by its spectral density $J(\omega) = \pi\sum_\nu \lambda_\nu^2
\delta(\omega-\omega_\nu)$ and the correlation function $C(t) =
\langle\xi(0)\xi(t)\rangle$.  At thermal equilibrium, the latter in
frequency space reads $C(\omega) = 2 J(\omega) n_\text{th}(\omega)$ with
the bosonic occupation number $n_\text{th}(\omega) =
[\exp(\omega/k_BT)-1]^{-1}$ and temperature $T$.  All numerical results
presented below are computed with an ohmic spectral density $J(\omega) =
\pi\alpha\omega/2$ with dimensionless coupling strength $\alpha$
\cite{LeggettRMP87}.  For the analytical arguments, however, it will only
be relevant that $J(\omega)$ smoothly vanishes for $\omega\to 0$.

Focussing on weak dissipation, we employ the Floquet-Bloch-Redfield
formalism of Refs.~\cite{KohlerPRE97, BlattmannPRA15} which provides a
master equation in Floquet basis of the form $\dot\rho_{\alpha\alpha'}(t) =
\sum_{\beta,\beta'} \mathcal{L}_{\alpha\alpha',\beta\beta'}(t)
\rho_{\beta\beta'}(t)$.  Generally, the Liouvillian $\mathcal{L}$ is
$2\pi/\Omega$-periodic, but within a rotating-wave approximation can be
replaced by its time-average.  While this formulation is suitable for
computational purposes, for our analytical treatment we employ a secular
approximation.  For $\alpha\lll 1$ the density operator eventually
becomes diagonal \cite{KohlerPRE97}, such that the stationary state can be
computed from the Pauli-type master equation $\dot p_\alpha = \sum_\beta
(\Gamma_{\alpha\beta}p_\beta - \Gamma_{\beta\alpha} p_\alpha)$ for the
Floquet-state populations $p_\alpha$ with the golden-rule rates
\cite{BlumelPRA91, KohlerPRE97, GrifoniPR98}
\begin{equation} \Gamma_{\alpha\beta}
= \sum_k C(q_\alpha-q_\beta+k\Omega) |S_{\alpha\beta,k}|^2 ,
\label{Gamma}
\end{equation}
where $S_{\alpha\beta,k}$ denotes the $k$th Fourier component of the
time-dependent transition matrix element $S_{\alpha\beta}(t) =
\langle\phi_\alpha(t)|S|\phi_\beta(t)\rangle$.  In Sambe formalism, it
can be written as $S_{\alpha\beta,k} =
\llangle\phi_\alpha|S|\phi_\beta^{(k)}\rrangle =
\llangle\phi_\beta|S|\phi_\alpha^{(-k)}\rrangle^*$, which allows an
elegant formulation of the symmetry properties.

Close to an intersection, the correlation function $C(\omega)$
in Eq.~\eqref{Gamma} is evaluated at the frequencies $\omega = \pm
r+k\Omega$, where the quasienergy splitting obeys $r\ll\Omega$.  Therefore,
$r$ can generally be neglected, such that the dissipative rates depend only
on the basis states $|\varphi_\pm\rrangle$ and the angle
$\vartheta$.  However, there exist also exceptional cases in which
$S_{\alpha\beta}(t)$ is practically time-independent, such that in
Eq.~\eqref{Gamma} the term with $k=0$ dominates.  Then
$\Gamma_{\alpha\beta}$ is formally given by the corresponding expression
for a time-independent problem, but with the eigenenergies replaced by
quasienergies, which is the constituting feature of a Floquet-Gibbs state
\cite{ShiraiPRE15, ShiraiNJP16}.

\begin{figure}
\centerline{\includegraphics{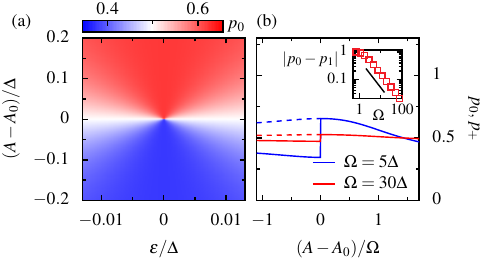}}
\caption{(a) Population $p_0$ of the Floquet state with lower quasienergy
as a function of the detuning $\varepsilon$ and the distance of the
amplitude to the crossing, $A-A_0$, where $A_0 \approx 2.4 \Omega$ is the
smallest driving amplitude at which the quasienergies cross for
$\Omega=5\Delta$.  The bath with temperature $T = 0.01\Delta/k_B$ is
coupled with strength $\alpha=0.01$ via $S=\sigma_x$.
(b) Population $p_0$ (solid line) for $\varepsilon=0$ exhibiting a
discontinuity at $A=A_0$ visible as step. The dashed line depicts the
population of the Floquet state with even generalized parity,
$|\phi_+\rrangle$, which is continuous.
Inset: Discontinuity of $p_0$ at the intersection as a function of the driving
frequency $\Omega$. The solid line indicates a decay $\propto 1/\Omega$.}
\label{fig:populationX}
\end{figure}

\subsection{The generic behavior}
We start with a bath coupling via $S=\sigma_x$ which represents the generic
case in which the Fourier coefficients $S_{\alpha\beta,k}$ with $k\neq 0$
dominate, such that $r$ can be neglected.  As already discussed, upon
$\vartheta \to \vartheta+\pi$ the Floquet states \eqref{phi0} and
\eqref{phi1} are interchanged.  Consequently, the transition rates
$\Gamma_{01}$ and $\Gamma_{10}$ and, thus, the populations $p_0$ and $p_1$
are interchanged as well, i.e., $p_0$ at an angle $\vartheta$ must be equal
to $p_1=1-p_0$ at $\vartheta+\pi$.  In short, the stationary populations
exhibit the duality of the Floquet states, such that $p_0-1/2$ turns out
anti-symmetric upon point reflection at the cone tip $(0,A_0)$, which is
confirmed by the numerical data shown in Fig.~\ref{fig:populationX}(a).
The same reasoning can be applied to the mean energy with the conclusion
that $E_0$ must change from $E_-$ for $\vartheta=0$ to $E_+$ for
$\vartheta=\pi$ in agreement with the result of the perturbation theory in
Eq.~\eqref{E0}.  Therefore the population difference $p_0-p_1$ and the mean
energy difference $E_1-E_0$ vary in the same way.
Interestingly, this includes a sign change of $p_0-p_1$ on a manifold with
constant quasienergy splitting.  On a line through the tip, this
leads to discontinuity which can be attributed to the discontinuities of
the Floquet states \eqref{phi0} and \eqref{phi1} \cite{EngelhardtPRL19}.
In a continuous basis such as the one formed by the states with definite
generalized parity along the $A$-axis, by contrast, the populations are
continuous as well.  This is verified by the dashed lines in
Fig.~\ref{fig:populationX}(b).

Since the populations as a function of $\vartheta$ change from $p_0<p_1$ to
the opposite order, there exists some values for which $p_0 = p_1 = 1/2$,
provided that the behavior is continuous.  Figure~\ref{fig:populationX}(a)
suggests, that this occurs at $\vartheta = \pm\pi/2$.  For a proof, we use
$G^\dagger\sigma_x G = \sigma_x$ together with Eqs.~\eqref{Gphi0} and
\eqref{Gphi1}.  For $\vartheta = \pm\pi/2$ this provides the relation
$S_{01,k} = \llangle\phi_0|G^\dagger\sigma_x G|\phi_1^{(k)}\rrangle =
(-1)^k S_{10,k}$.  Therefore, all terms that contribute to the transition
rate $\Gamma_{01}$ also occur in the expression for $\Gamma_{10}$ such that
$\Gamma_{01} = \Gamma_{10}$ and, thus, $p_0 = p_1$.  In other words, the
stationary density operator is proportional to the unit matrix, i.e., it
describes a fully incoherent mixture irrespective of the temperature.
Such equal population of both Floquet states is particularly interesting,
because it can be detected by dispersive readout, i.e., it leaves its
fingerprints in the transmission of a cavity coupled to the two-level
system, as has been demonstrated in a recent experiment~\cite{ChenPRB21}.

The symmetry properties discussed so far hold for any driving frequency
$\Omega\gtrsim\Delta$. Within the usual high-frequency approximation
\cite{GrossmannEL92} (for a summary, see the Appendix), we find
a further symmetry relation.  It is based on the transformation with
$\sigma_z$, which in the Hamiltonian \eqref{H} inverts the
sign of the tunnel term, effectively causing $\Delta\to -\Delta$.
Since for large frequencies the quasienergies read $q_{0,1} = \mp(\Delta/2)
J_0(A/\Omega)$ \cite{GrossmannEL92}, the state
$\sigma_z|\phi_0(t)\rangle$ must be proportional to the Floquet state with
opposite quasienergy, i.e., $\sigma_z|\phi_0(t)\rangle \propto
|\phi_1(t)\rangle$ and vice versa.  Thus, for any value of $\vartheta$,
$\sigma_z$ acts on the Floquet states like the generalized parity operator
$G$ for the particular value $\vartheta=\pi/2$.  Therefore, we expect to
find in the high-frequency limit the same consequences, but now in the
whole vicinity of the intersection.  Specifically, for any $\vartheta$, the
stationary state is fully mixed state with $p_0 = p_1 = 1/2$.

It is interesting to see how finite-frequency corrections vanish with
increasing $\Omega$.  The inset of Fig.~\ref{fig:populationX}(b) reveals
that the limit is approached $\propto 1/\Omega$.  Even for
$\Omega=15\Delta$, the difference $\sim0.1$ is still remarkably large.
Thus, while the usual high-frequency approximation for the quasienergies is
reliable for $\Omega \gtrsim 2\Delta$ \cite{GrossmannEL92}, we observe that
the approximation describes dissipative effects well only for larger
frequencies.

\begin{figure}
\centerline{\includegraphics{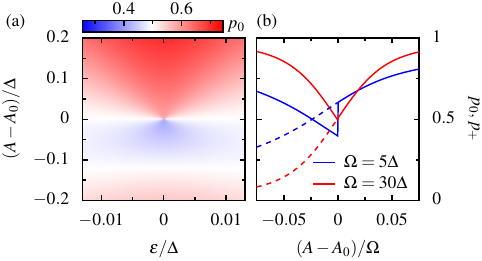}}
\caption{The same as Fig.~\ref{fig:populationX} but for a bath coupling via
$S=\sigma_z$.}
\label{fig:populationZ}
\end{figure}

\subsection{Coupling via $\sigma_z$ and Floquet-Gibbs state}

A bath coupling via $S=\sigma_z$ anti-commutes with the generalized parity
operator and, thus, $G^\dagger\sigma_z G = -\sigma_z$.  Then the symmetry
relation for the matrix element of $S$ acquires a minus sign such that
$S_{01,k} = (-1)^{k+1} S_{10,k}$.  Since the rates $\Gamma_{\alpha\beta}$
depend only on absolute values, the generic behavior discussed above should
be found as well.  However, as we can appreciate in
Fig.~\ref{fig:populationZ}(a), this is the case only very close to the
intersection, while for a slightly smaller amplitude, the population of
$\phi_0$ becomes much larger.  The reason for this is that there the
Fourier component with $k=0$ dominates.  This becomes clear when computing
the matrix elements within the usual high-frequency approximation,
$\langle\phi_0(t)|\sigma_z|\phi_1(t)\rangle= -1$, which can be easily
verified by the explicit expressions given in the Appendix.  Hence,
$S_{10}(t)$ is time independent such that $S_{\alpha\beta,k}$ vanishes
unless $k=0$.  Therefore, as discussed above, we expect to find a
Floquet-Gibbs state.  Indeed the conditions for the emergence of such state
formulated in Refs.~\cite{ShiraiPRE15, ShiraiNJP16} are fulfilled, namely
the driving frequency is very large, while the driving and the system-bath
coupling commute at all times.

We are also interested in intermediate frequencies for which components
with $k\neq0$ still contribute.  Very close to the intersection, $r$ is
sufficiently small such that the term with $k=0$ may be neglected.  Then the
population $p_0$ still resembles the generic case, including the
discontinuity at $A_0$.  When moving away from the
intersection, however, $r$ grows such that the contribution of
$S_{\alpha\beta,0}$ becomes increasingly important and one observes a
crossover from the generic behavior to a Floquet-Gibbs state.  Data along
the $A$-axis in Fig.~\ref{fig:populationZ}(b) verify that for
$\Omega=5\Delta$, features of the generic behavior are still present.  With
increasing frequency, the crossover region shrinks and eventually
disappears.

As mentioned above, equal Floquet state population can be detected by
coupling the driven system to a cavity \cite{ChenPRB21}.  Therefore, the
crossover to a Floquet-Gibbs state may be tested experimentally as well.
Such measurement may provide evidence for a system-bath coupling via
$\sigma_z$.

\subsection{Mixed bath coupling}

\begin{figure*}
\centerline{\includegraphics{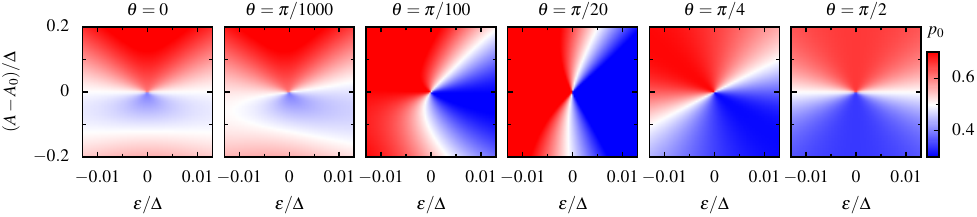}}
\caption{Population of the Floquet state $\phi_0$ for system-bath
coupling via the operator $S$ defined in Eq.~\eqref{app:sigmatheta}
with the angle $\theta$ given on the top of each panel. The driving
frequency is $\Omega=8\Delta$.}
\label{fig:mixed}
\end{figure*}

One might suppose that at least for moderate driving frequency, a bath
coupling via any linear combination of Pauli matrices $S=\sum_i
c_i\sigma_i$ would also lead to the generic behavior depicted in
Fig.~\ref{fig:populationX}(a).  Now, by contrast,
the condition of $S$ being either even or odd under generalized parity no
longer holds.  Since $\sigma_x$ is even while $\sigma_y$ and $\sigma_z$ are
odd, the requirement is met only if either $c_1=0$ or $c_2=c_3=0$. For a
visualization of the crossover, we consider a bath coupling via
\begin{equation}
S = \sigma_x\sin\theta + \sigma_z\cos\theta .
\label{app:sigmatheta}
\end{equation}
Figure \ref{fig:mixed} shows the population $p_0$ for various values of the
coupling angle $\theta$.  In consistency with Refs.~\cite{BlattmannPRA15,
EngelhardtPRL19}, it shows that the coupling via $\sigma_z$ is rather
sensitive to any small admixture of $\sigma_x$.  Already for a rather small
value of $\theta$, the Floquet-Gibbs state is lost.  With increasing mixing
angle $\theta$, the line with equal population $p_0=p_1=1/2$ turns smoothly
into the one for pure $\sigma_x$ coupling.

\begin{figure}
\centerline{\includegraphics{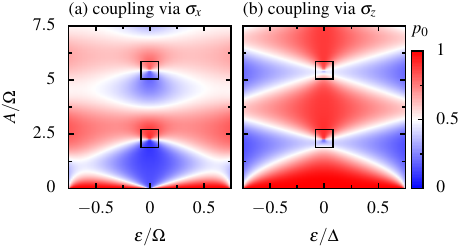}}
\caption{Floquet state populations for bath coupling via $\sigma_x$ (a) and
$\sigma_z$ (b), respectively, cf.\ Figs.~\ref{fig:populationX}(a) and
\ref{fig:populationZ}(a).  The driving frequency $\Omega=1.5\Delta$ is
chosen rather low, such that in panel (b) the zone with deviations from the
Floquet-Gibbs state at $\varepsilon=0$ and $A\lesssim A_0$ becomes visible.
All other parameters are as in Fig.~\ref{fig:populationX}.
The rectangles mark the conical intersections.}
\label{fig:global}
\end{figure}

\subsection{Global picture}

While we have worked out the scenario in the vicinities of conical
intersections, it remains to integrate these findings into a global picture.
Therefore, we show in Fig.~\ref{fig:global} the population in a much larger
parameter range.  The stripes with equal population (white) indicate that
the choice of $S$ influences the global behavior significantly.
For a coupling via $\sigma_x$, they are merely bent, but
nevertheless extend over a rather large range.  For $\sigma_z$, by
contrast, the switching between the generic behavior and the Floquet-Gibbs
state brings about a further line with $p_0=1/2$.

\section{Conclusions}
\label{sec:conclusions}

We have worked out features of the driven two-level system in the
vicinity of conical intersections of quasienergies, which can be derived
from the behavior under transformation with the generalized parity
operator.  In the absence of a detuning, the according symmetry is
responsible for the emergence of exact crossings.  For a small detuning,
i.e., when the Hamiltonian no longer obeys generalized parity, we found
various generic properties of the Floquet states, foremost, the behavior of
the mean energies and of the dissipative transition rates.
The analytical achievements are based on a rotating-wave approximation
which, as our numerical results verify, holds even for relatively small
driving frequencies of the order of the tunnel matrix element.  In
particular, we did not employ the common high-frequency approximation with
Bessel functions.

While the quasienergies form a double cone, the mean energies exhibit a
more intricate structure.  On manifolds with constant quasienergies, they
are continuously interchanged.  Generally, the same happens for the
populations of the Floquet states in the stationary limit. Therefore, the
mean energies rather than the quasienergies provide an indication for the
populations.  Moreover, the sign change of the population difference
implies the existence of fully mixed stationary states characterized by
equal population of both Floquet states.  In the high-frequency limit, such
maximal entropy states are found in the whole vicinity of the intersection,
which follows from a hidden chirality.
An exceptional case is realized when the system-bath coupling commutes with
the driving.  Then the generic behavior of the populations, including the
discontinuity along the amplitude axis, is found only for intermediate
driving frequencies in a limited region.  With increasing frequency,
this region shrinks and eventually a Floquet-Gibbs state is formed.

At a larger distance from the cone, we find that the shape of the manifolds
with maximal entropy depends on the system-bath interaction operator.  By
coupling in addition the two-level system to a microwave cavity, one can in
principle find experimental signatures for equal Floquet state population
\cite{ChenPRB21}.  Such measurements may provide evidence for the
crossover from the generic behavior governed by mean energies to a
Floquet-Gibbs state and, hence, for the type of bath coupling.

\begin{acknowledgments}
This work was supported by the Spanish Ministry of Science, Innovation, and
Universities (Grant No.\ PID2020-117787GB-I00), and by the CSIC Research
Platform on Quantum Technologies PTI-001.
\end{acknowledgments}

\appendix

\begin{figure}
\centerline{\includegraphics{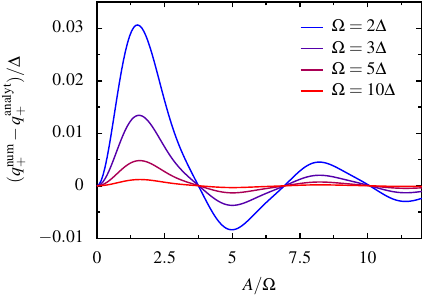}}
\caption{Difference of the numerically computed quasienergy and the
analytical approximation in Eq.~\eqref{app:q} as a function of the driving
amplitude.}
\label{fig:bessel}
\end{figure}

\section{High-frequency approximation}
\label{app:hf}

For completeness and reference, we provide the known expressions
for the Floquet states in the limit of large driving frequencies
\cite{GrossmannEL92},
\begin{equation}
|\phi_\pm(t)\rangle = \frac{1}{\sqrt{2}}
\begin{pmatrix} \pm e^{-iF(t)} \\ e^{iF(t)} \end{pmatrix},
\label{app:phi}
\end{equation}
with $F(t) = (A/2\Omega)\sin(\Omega t)$ and their quasienergies
\begin{equation}
q_\pm = \pm \frac{\Delta}{2} J_0(A/\Omega)
\label{app:q}
\end{equation}
with $J_0$ the zeroth-order Bessel function of the first kind.  Thus, the
position $A_0$ of any conical intersection requires that $A_0/\Omega$
matches a root of $J_0$.

For a derivation of Eqs.~\eqref{app:phi} and \eqref{app:q}
\cite{GrossmannEL92}, one uses the Hamiltonian in the absence of the tunnel
term as zeroth order, $H_0(t) = (A/2)\sigma_z \cos(\Omega t)$, where the
corresponding propagator reads $U_0(t) = \exp[-i(A/2\Omega)\sigma_x
\sin(\Omega t)]$.  Transformation of the remaining contribution, $H_1 =
(\Delta/2)\sigma_z$ with $U_0$ and subsequent time-average, i.e.\
a rotating-wave approximation, yields $H_1' =
(\Delta/2)J_0(A/\Omega)\sigma_x$.  Its eigenvalues are the quasienergies
\eqref{app:q},
while the Floquet states in Eq.~\eqref{app:phi} are the eigenstates of $H'$
transformed back to the Schr\"odinger picture.  Figure \ref{fig:bessel}
shows that already for $\Omega\gtrsim 2\Delta$, the approximation in
Eq.~\eqref{app:q} deviates from the exact value by only a few percent of
the tunnel matrix element $\Delta$.

A particular feature of the Floquet states in Eq.~\eqref{app:phi} is that
the transition matrix element $\langle\phi_+(t)|\sigma_z|\phi_-(t)\rangle =
-1$ is time-independent.  For $\vartheta=0$, we have $\phi_0 = \phi_-$ and
$\phi_1=\phi_+$ (and vice versa for $\vartheta=\pi$) such that only the
Fourier component with $k=0$ contributes to the dissipative rates
$\Gamma_{01}$ and $\Gamma_{10}$.  For sufficiently large driving frequency, this
finally leads to the emergence of a Floquet-Gibbs state \cite{ShiraiPRE15,
ShiraiNJP16}. For intermediate $\Omega$, however, components with $k\neq 0$
still play a role such that sufficiently close to $A_0$, the generic
behavior is restored.

Within high-frequency approximation, the Floquet states possess
a further symmetry, namely a hidden chirality given by a transformation
with $\sigma_z$.  Obviously, in the Hamiltonian in Eq.~\eqref{H} of the
main text, the terms with
the detunings $\varepsilon$ and $A\cos(\Omega t)$ are invariant under
transformation with $\sigma_z$.  By contrast, the tunnel term changes sign,
effectively causing $\Delta\to -\Delta$.  Thus, already from
Eq.~\eqref{app:q} it becomes clear that $\sigma_z|\phi_0(t)\rangle$ and
$|\phi_1(t)\rangle$ can differ at most by a phase factor.  For the
choice of the phases of the Floquet states in Eq.~\eqref{app:phi}, one finds
\begin{equation}
\sigma_z|\phi_\pm(t)\rangle = -|\phi_\mp(t)\rangle .
\end{equation}
Relations of this kind are typically found as consequence of a chirality,
i.e., a mapping that inverts the sign of the Hamiltonian.  Notice that
$\sigma_z$ and, thus, the relation $\sigma_z H_1' \sigma_z = -H_1'$ are not
affected by the back transformation to the Schr\"odinger picture.

\end{document}